\documentclass[mathleft
]{an}
\usepackage{graphicx}
\usepackage{times}
\begin{document}

\Pagespan{1}{}
\Yearpublication{2014}%
\Yearsubmission{2014}%
\Month{8}%
\Volume{334}%
\Issue{8}%
\DOI{This.is/not.aDOI}%

\title{Classical Cepheids in the  Galactic outer ring $R_1R_2'$}
\author{A.~M. Mel'nik\inst{1}\thanks{Corresponding author.
\email{anna@sai.msu.ru}}, P. Rautiainen\inst{2}, L.~N.
Berdnikov\inst{1,3}, A.~K. Dambis\inst{1} \& A.~S.
Rastorguev\inst{1}}
\authorrunning{A.M. Mel'nik et al.}
\titlerunning{Classical Cepheids in the  Galactic outer ring $R_1R_2'$}

\institute{Sternberg Astronomical Institute, Lomonosov Moscow
State University, 13 Universitetskij Prosp., Moscow 119991,
Russia \and Department of Physics/Astronomy Division, University
of Oulu,  P.O. Box 3000, FIN-90014 Oulun yliopisto, Finland \and
Astronomy and Astrophysics Research division, Entoto Observatory
and Research Center, P.O.Box 8412, Addis Ababa, Ethiopia}

\received{2014} \accepted{2014} \publonline{2014}


\abstract{The kinematics and distribution of classical Cepheids
within $\sim3$ kpc from the Sun suggest the existence of the
outer ring $R_1R_2'$ in the Galaxy. The optimum value of the
solar position angle with respect to the major axis of the bar,
$\theta_b$, providing the best agreement between the distribution
of Cepheids and model particles  is $\theta_b=37\pm13^\circ$. The
kinematical features obtained for Cepheids with negative
Galactocentric radial velocity $V_R$ are consistent with the
solar location near the descending segment of the outer ring
$R_2$. The sharp rise of extinction toward of the Galactic center
can be explained by the presence of the outer ring $R_1$ near the
Sun.}

\maketitle


\newpage
\begin{figure*} \centering
\resizebox{\hsize}{!}{\includegraphics{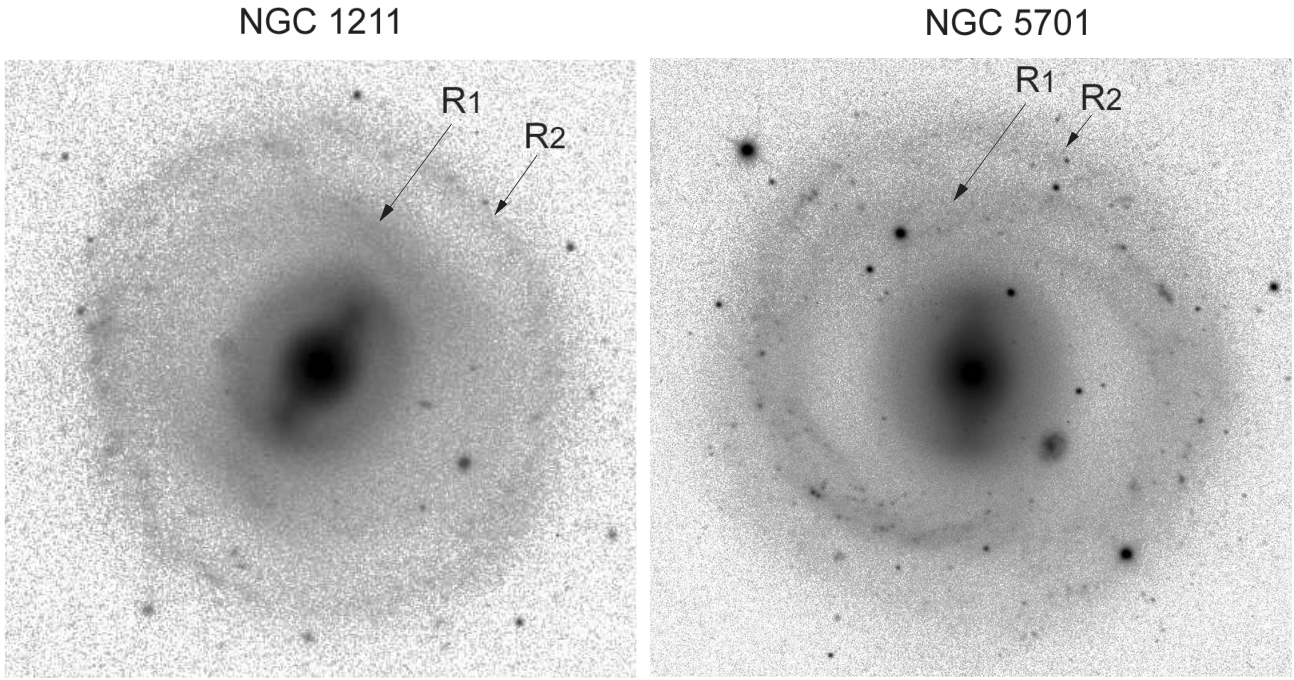}} \caption{
Galaxies NGC 1211 and NGC 5701 with outer ring morphology
$R_1R_2$.  The g-band images  were taken from NASA-Sloan Atlas
(http://www.nsatlas.org/) created by  Michael Blanton et al. (see
also Blanton et al. 2011). The images were obtained in green
filter centered at 477 nm during the program of the Sloan Digital
Sky Survey (York et al., 2000). The images demonstrate two outer
rings oriented perpendicular to each other. Of the two  rings,
$R_1$ is located a bit closer to the galactic center than the
$R_2$.} \label{rings}
\end{figure*}
\newpage
\begin{figure*} \centering
\resizebox{12 cm}{!}{\includegraphics{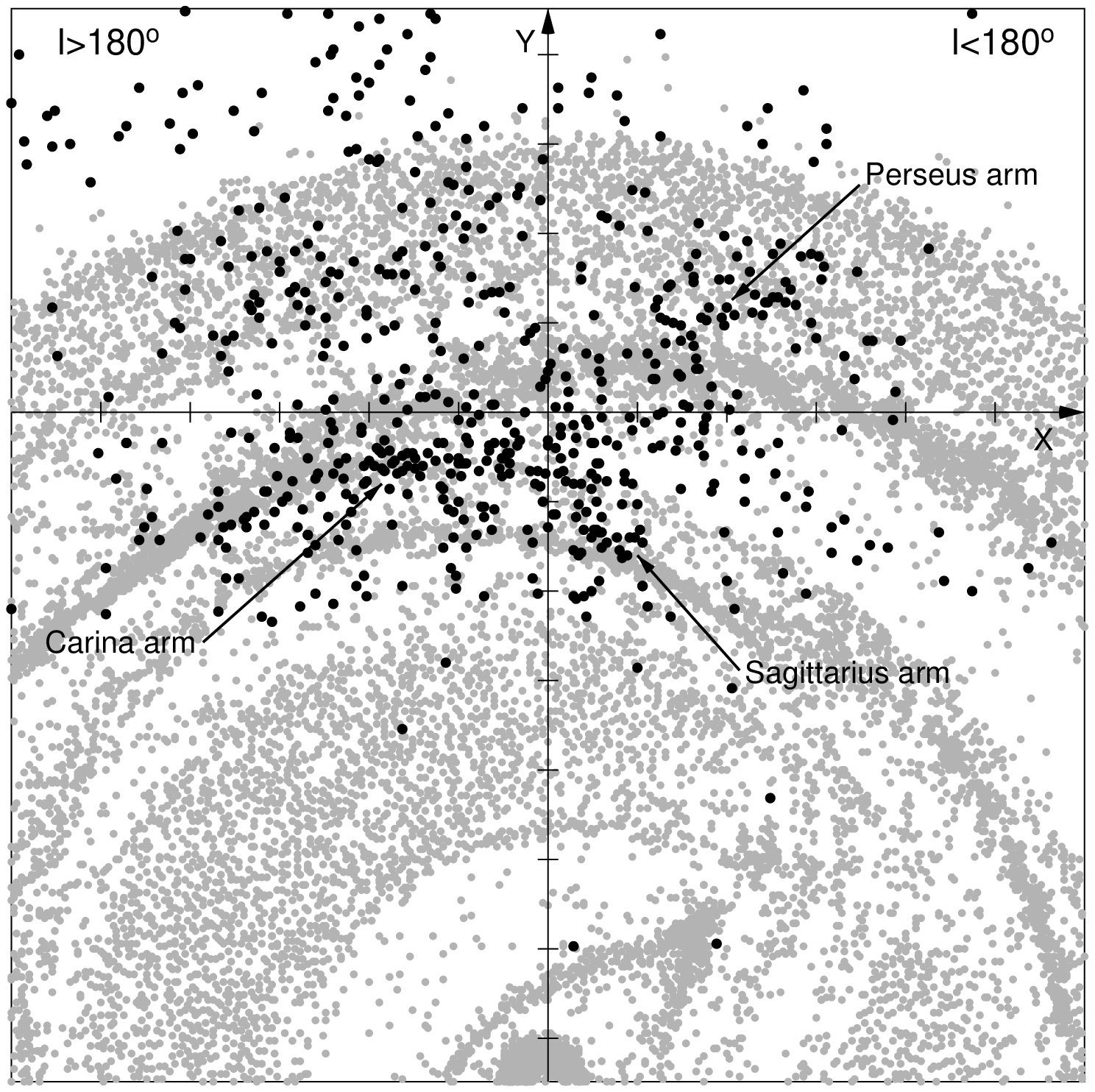}} \caption{
Distribution of classical Cepheids (black circles) and model
particles (gray circles) in the Galactic plane.  For model
particles the solar position angle is chosen to be
$\theta_b=45^\circ$. The arrows indicate the positions of the
Sagittarius, Carina, and  Perseus arm-fragments. The $X$-axis
points in the direction of Galactic rotation and the $Y$-axis is
directed away from the Galactic center. One tick interval along
the $X$- and $Y$-axis corresponds to 1 kpc. The Sun is at the
origin.  We can see a fork-like structure in the distribution of
Cepheids: on the left  ($l>180^\circ$) Cepheids strongly
concentrate to the only one arm (the Carina arm), while on the
right ($l<180^\circ$) there are two arm-fragments located near
the Perseus and Sagittarius regions.} \label{distrib}
\end{figure*}

\section{Introduction}

Classical Cepheids -- F--K-type supergiants with ages from 20 to
200 Myr (Efremov  2003, Bono et al. 2005) -- are good tracers of
the Galactic spiral structure and regions of high gas density.
Due to the period-luminosity relation the distances to classical
Cepheids can be determined with an accuracy of $\sim10$\%
(Berdnikov, Dambis \& Vozyakova 2000; Sandage \& Tammann 2006).

Most of the recent  studies of the spiral structure of the Galaxy
(see e.g.\ the review by Vall{\'e}e (2013) and references
therein) have typically suggested  a 2- or 4-armed spiral pattern
with a pitch angle of nearly $i=6^\circ$ or 12$^\circ$,
respectively. In our previous studies (Mel'nik \& Rautiainen
2009; Rautiainen \& Mel'nik 2010) we proposed an alternative to
these purely spiral models -- a two-component outer ring
$R_1R_2'$.

The main advantage of the 4-armed  spiral model is that it can
explain the distribution of HII regions in the Galactic disk
(Georgelin \& Georgelin 1976; Russeil 2003; and other papers) and
the existence of so-called tangential directions related to the
maxima in the thermal radio continuum, HI and CO emission which
are associated with the tangents to the spiral arms (Englmaier \&
Gerhard 1999; Vall{\'e}e 2008).

The main shortcoming of this approach is the absence of a
dynamical mechanism to maintain the   global spiral pattern for a
long time period (more than a few  disk rotations) (Toomre 1977;
Athanassoula 1984). Generally, the bar could support the spiral
pattern,  but in this case  the spiral pattern must rotate with
the angular velocity of the bar ($\Omega_s=\Omega_b$) (Englmaier
\& Gerhard 1999). The estimates of the corotation radius (CR) of
the Galactic bar do not exceed 5 kpc, which gives the lower limit
for its angular velocity $\Omega_b>40$ km s$^{-1}$ kpc$^{-1}$
(Weiner \& Sellwood 1999). However, this model cannot explain the
kinematics of the Perseus region: the direction of the velocities
of young stars in the Perseus region, if interpreted in terms of
the density-wave concept (Lin, Yuan  \& Shu 1969), indicates that
a fragment of a Perseus arm must be located inside the corotation
circle (CR) (Burton \& Bania 1974; Mel'nik, Dambis \& Rastorguev
2001; Mel'nik 2003; Sitnik 2003) implying an upper limit
$\Omega_s<25$ km s$^{-1}$ kpc$^{-1}$ for its pattern speed, which
is inconsistent with the one mentioned above. Attempts have been
undertaken to overcome this contradiction by introducing an
analytical spiral potential rotating slower than the bar
(Bissantz, Englmaier  \& Gerhard 2003). Note that numerical
simulations show that galactic stellar disks can develop modes
that rotate slower than the bar (Sellwood \& Sparke 1988; Masset
\& Tagger 1997; Rautiainen \& Salo 1999, 2000). However, it is
questionable whether the amplitude of slow modes can be large
enough to determine the kinematics at twice the radius of the CR
of the Galactic bar.

Another concept of  the Galactic spiral structure is that the
disk rather than global modes forms  transient spiral arms.
Sellwood (2000, 2010) advances the idea that fresh and decaying
instabilities are connected through  resonances, which is based
on some specific features in the angular momentum distribution of
old stars in the solar vicinity. Baba et al.~(2009) build a model
of the Galaxy with transient spiral arms, which can explain the
large peculiar velocities, 20--30 km s$^{-1}$, of maser sources
in the Galactic disk. Transient spiral arms must heat the
galactic disk in few disk rotation periods (Sellwood \& Carlberg
1984). We can suggest that young massive stars born in such arms
should also acquire large peculiar velocities. However, young
stellar objects (classical Cepheids, young open clusters,
OB-associations) in the wide solar neighborhood move in nearly
circular orbits with  average velocity deviations of 7--13 km
s$^{-1}$ from the rotation curve (Zabolotskikh, Rastorguev \&
Dambis 2002; Mel'nik \& Dambis 2009; Bobylev \& Baikova 2012; and
other papers).

Models of the Galaxy with the outer ring $R_1R_2'$  can reproduce
well the radial and azimuthal components of the residual
velocities (after substracting the velocity due to the rotation
curve and the solar motion to the apex) of OB-associations in the
Sagittarius and Perseus stellar-gas complexes identified by
Efremov \& Sitnik (1988). The radial velocities  of most
OB-associations in the Perseus region are directed toward the
Galactic center and this indicates the presence of the  ring
$R_2$ in the Galaxy, while the radial velocities in the
Sagittarius region are directed away from the Galactic center
suggesting the existence of the ring $R_1$. The nearly zero
azimuthal component of the residual velocity of most
OB-associations in the Sagittarius region precisely constrains
the solar position angle with respect to the bar major axis,
$\theta_b=45\pm5^\circ$. We considered models with analytical
bars and N-body simulations (Mel'nik \& Rautiainen 2009;
Rautiainen \& Mel'nik 2010).

Models of a two-component  outer ring are consistent with the (l,
$V_{LSR}$) diagram by Dame, Hartmann \& Thaddeus (2001). These
models can explain the position of the Carina arm with respect to
the Sun and with respect to the bar. They can also explain the
existence of some of the tangential directions corresponding to
the emission maxima near the terminal velocity curves which, in
this case, can be associated with the tangents to the outer and
inner rings. Our model diagrams (l, $V_{LSR}$) reproduce the
maxima in the direction of the Carina, Crux (Centaurus), Norma,
and Sagittarius arms. Additionally, N-body model yields  maxima
in the directions of the Scutum and 3-kpc arms  (Mel'nik \&
Rautiainen 2011,  2013).

The elliptic outer rings can be divided into the ascending and
descending segments: in the ascending segments  Galactocentric
distance $R$ decreases with  increasing azimuthal angle $\theta$,
which itself increases in the direction of Galactic rotation,
whereas the dependence is reversed in the descending segments.
Ascending and descending segments of the rings can be regarded as
fragments of  trailing and leading spiral arms, respectively.
Note that  if considered  as fragments of the spiral arms, the
ascending segments of the outer ring $R_2$ have the pitch angle
of $\sim6^\circ$ (Mel'nik \& Rautiainen 2011).

Two main classes of  outer rings and pseudorings (incomplete
rings made up of two tightly wound spiral arms) have been
identified: rings $R_1$ (pseudorings $R'_1$) elongated
perpendicular to the bar and   rings $R_2$ (pseudorings $R'_2$)
elongated parallel to the bar.  In addition, there is a combined
morphological type $R_1R_2'$ which exhibits elements of both
classes (Buta 1995; Buta \& Combes 1996; Buta \& Crocker 1991;
Comeron et al.~2013).  Modelling shows that outer rings are
usually located near the OLR of the bar (Schwarz 1981; Byrd et
al.~1994; Rautiainen \& Salo 1999, 2000; and other papers).
However,  the ring $R_1$ can be located further inwards, closer
to the outer 4/1-resonance in some cases (Treuthardt et
al.~2008).

Buta and Combes (1996) have shown that for the galaxies in the
lower red-shift range the frequency of the outer rings is 10\% of
all types of spiral galaxies. However, it increases to 20\% for
the early-type sample. Note, however, that  Comeron et al. (2014)
obtained a larger frequency from an analysis of the data from
mid-infrared survey (Spitzer Survey of Stellar Structure in
Galaxies, Sheth at al. 2010): 16\% for all spiral galaxies
located inside 20 Mpc and over 40\% for disk galaxies of early
morphological types (galaxies with large bulges). They have also
found that the frequency of outer rings increases from $15\pm
2$\% to $32\pm 7$\% when going through the family sequence from
SA to SAB, and decreases again to $20\pm2$\% for SB galaxies. The
catalog of Southern Ringed Galaxies by Buta (1995) gives the
following statistics of the main ring classes: 18\% ($R_1$), 37\%
($R_1'$), under 1\% ($R_2$), 35\% ($R_2'$), and 9\% ($R_1R_2'$).
Comeron et al. (2014)  generally confirm these results:  $35\pm
18$\% of outer rings in barred galaxies are parallel to the bar
and $65\pm 39$\% are oriented perpendicular to it. The small
fraction of galaxies with $R_1R_2'$ rings may be due to the
selection effects -- the rings $R_2$ are often weaker than rings
$R_1$, moreover rings $R_2$ are often appear more conspicuous
when observed  in the B-band. The completeness of the catalogs
regarding  the $R_2$ features is difficult to estimate.
Generally, the frequency of galaxies with outer rings $R_1R_2'$
among barred galaxies may be as high as few per cent. Note that
the catalog by Buta (1995) includes  several tens of galaxies
with rings $R_1R_2'$.

Figure~\ref{rings} shows  two galaxies with $R_1R_2$ outer ring
morphology. The g-band images of galaxies NGC 1211 and NGC 5701
were taken from NASA-Sloan Atlas (http://www.nsatlas.org/)
created by  Michael Blanton et al. (see also Blanton et al.
2011). The images were obtained in green filter centered at 477
nm during the program of the Sloan Digital Sky Survey (York et
al., 2000). The images demonstrate two outer rings oriented
perpendicular to each other. Of the two  rings, $R_1$ is located
a bit closer to the galactic center than the $R_2$. Other
examples of galaxies with the $R_1R_2'$ morphology that can also
be viewed as possible prototypes of the Galaxy are ESO 245-1, NGC
1079,  NGC 3081, NGC 5101, NGC 6782, and NGC 7098. Their images
can be found in de Vaucouleurs Atlas of Galaxies by Buta, Corwin,
Odewahn (2007) at http://bama.ua.edu/$\sim$rbuta/devatlas/

Schwarz (1981)  associates   two main types of  outer rings with
two main families of periodic orbits existing near the OLR of the
bar (Contopoulos \& Papayannopoulos 1980). The main periodic
orbits $x_1(1)$ and $x_1(2)$ (in terms of the nomenclature of
Contopoulos \& Grosbol 1989) are followed by numerous chaotic
orbits, and this guidance enables elliptical rings to hold a lot
of gas in their vicinity. The rings $R_1$ are supported by
$x_1(2)$-orbits  lying inside the OLR and elongated perpendicular
to the bar while the rings $R_2$ are supported by $x_1(1)$-orbits
located slightly outside the OLR and elongated along the bar.
However, the role of chaotic and periodic orbits appears to be
different in the center region and on the galactic periphery:
chaos is dominant outside the CR, while most orbits in the bar
are ordered (Contopoulos \& Patsis 2006; Voglis, Harsoula \&
Contopoulos 2007; Harsoula \& Kalapotharakos 2009). Probably, it
is not only periodic orbits associated with the OLR that are
responsible for the formation of the outer rings/pseudorings. A
concept has been proposed that the formation of outer rings as
well as that of spiral arms is determined by manifolds associated
with the Lagrangian points L1 and L2 (Romero-G{\'o}mez et
al.~2007; Athanassoula et al.~2010).

The existence of the bar  in the Galaxy is confirmed by numerous
infra-red observations (Blitz \& Spergel 1991; Benjamin et
al.~2005; Cabrera-Lavers et al.~2007; Churchwell et al.~2009,
Gonz\'alez-Fern\'andez et al.~2012) and by gas kinematics in the
central region (Binney et al.~1991; Englmaier \& Gerhard 1999;
Weiner \& Sellwood 1999).  The general consensus is that  the
major axis of the bar is oriented in the direction
$\theta_b=15\textrm{--}45^\circ$ in such a way that the end of
the bar closest to the Sun lies in  quadrant I. The semi-major
axis of the Galactic bar is supposed to lie in the range
$a=3.5\textrm{--}5.0$ kpc. Assuming that its end is located close
to its corotation radius, i.e. we are dealing with a so-called
fast bar (Debattista \& Sellwood 2000), and that the rotation
curve is flat, we can estimate the bar angular speed $\Omega_b$
which appears to be constrained to the interval
$\Omega_b=40\textrm{--}65$ km s$^{-1}$ kpc$^{-1}$. This means
that the OLR of the bar is located in the solar vicinity:
$|R_{OLR}-R_0|<1.5$ kpc. Studies of the kinematics of old disk
stars in the nearest solar neighborhood, $r<250$ pc, reveal the
bimodal structure of the distribution of ($u$, $v$) velocities
which is also interpreted as a result of the solar location near
the OLR of the bar (Kalnajs 1991; Dehnen 2000; Fux 2001; and
other papers).

In this paper we  show that the morphological and kinematical
features of the Cepheid sample considered are consistent with the
presence of a ring $R_1R_2'$ in the Galaxy. Section 2 describes
the models and catalogues used; Section 3 considers the special
features in the  distribution of Cepheids; Section 4 studies the
kinematics of Cepheids, and Section 5 presents the main results
and their discussion.

\section{Models, catalogues, and  calibrations}

We used the  catalogue of classical Cepheids by Berdnikov et
al.~(2000) which is continuously improved and updated by
incorporating new observations (Berdnikov et al.~2009a, 2009b,
2011, 2014). The last version of the catalogue includes the data
for 674 Cepheids (Berdnikov, Dambis \& Vozyakova 2014, in
preparation). The procedure of deriving distances  is based on
the K-band period-luminosity relation of Berdnikov, Vozyakova \&
Dambis (1996b) and interstellar-extinction law derived in
Berdnikov, Vozyakova \& Dambis (1996a). The
interstellar-extinction values are estimated using the $B-V$
period-color relation of Dean, Warren, and Cousins (1978). Note
the natural spread of the period-color relation does not
introduce any substantial errors in the inferred distance values
because the $K$-band extinction is very small,
$A_K$~=~0.274$E_{B-V}$. Our procedure in this case is essentially
equivalent to using the $Vm_{\lambda}$ Wesenheit function with
the deviations from the mean period-luminosity and period-color
relations virtually cancelling each other (Berdnikov et al.
1996a).  Note the variations in the distance scale up to 10\% do
not affect our conclusions.

We use the simulation code developed by H. Salo (Salo 1991; Salo
\& Laurikainen 2000)  to construct two different types of models
(models with analytical bars and N-body simulations) which
reproduce the kinematics of OB-associations in the Perseus and
Sagittarius regions. Among many models with outer rings, we chose
model 3 from the series of models with analytical bars (Mel'nik
\& Rautiainen 2009) for comparison with observations. This model
has nearly flat rotation curve. The bar semi-axes are equal to
$a=4.0$~kpc and $b=1.3$~kpc. The positions and velocities  of
$5\times 10^4$ model particles (gas+OB) are considered at time
$T=15$ ($\sim$1 Gyr).  We scaled and turned this model with
respect to the Sun to achieve the best agreement between  the
velocities of model particles and those of OB-associations in
five stellar-gas complexes (Efremov \& Sitnik 1988).

We adopt a solar Galactocentric distance of $R_0=7.5$ kpc
(Rastorguev et al.~1994; Dambis, Mel'nik \& Rastorguev 1995;
Glushkova et al.~1998; Nikiforov 2004;  Feast et al.~2008;
Groenewegen, Udalski \& Bono 2008; Reid et al.~2009b; Dambis et
al.~2013). As model 3 was adjusted for $R_0=7.1$ kpc, we rescaled
all distances for model particles by a factor of $k=7.5/7.1$.
Note that the analysis of morphology and kinematics of stars
located within 3 kpc from the Sun is practically independent of
the choice of $R_0$ in the range 7--9 kpc.

Table 1, which is available in the online version of the paper,
gives the positions, line-of-sight velocities and proper motions
of classical Cepheids. It ia also available at:

http://lnfm1.sai.msu.ru/$\sim$anna/tables/tables\verb"_"2014.zip

It includes 674 classical Cepheids from the new release of the
catalog of classical Cepheids by Berdnikov et al. (2014, in
preparation). For every classical Cepheid we give its designation
in the General Catalog of Variable Stars (GCVS) (Samus at al.
2007) or  in the All Sky Automated Survey (ASAS) (Pojmanski
2002), its type (see GCVS description), fundamental period $P_F$,
intensity-mean V-band magnitude, J2000 equatorial coordinates
$\alpha$ and $\delta$, galactic coordinates $l$ and $b$, and
heliocentric distance $r$. Table~1 also gives the Cepheid
line-of-sight velocities $V_r$ (the so-called
$\gamma$-velocities, see Metzger, Caldwell \& Schechter 1992),
their uncertainties $\varepsilon_{Vr}$ and the references (1--6)
to the sources from which they are taken. The proper motions of
Cepheids were adopted from the new reduction of Hipparcos data
(ESA 1997) by van Leeuwen (2007). Table~1 presents proper motions
$\mu_{\alpha}$ and $\mu_{\delta}$, their uncertainties
$\varepsilon_{\mu\alpha}$ and $\varepsilon_{\mu\delta}$ and the
corresponding  Hipparcos catalog number $n_{Hip}$.

Table~2 lists the catalogs of  line-of-sight velocities used in
our study.  For each catalog it gives the corresponding reference
number used in Table 1, the authors of the catalog, the full
reference to the paper/catalog, and the number of velocities
taken from each source. Some sources include a series of papers
by one group of researches and/or data available only online.

\addtocounter{table}{1}

\begin{table*}
 \caption{\bf Sources of  Cepheid line-of-sight velocities}
 \begin{tabular}{llccc}
    \\[-7pt] \hline\\[-7pt]
 Ref.& Authors & year & journal  & number\\
  \\[-7pt] \hline\\[-7pt]
 &Metzger et al. & 1991 & ApJS, 76, 803  &\\
 &Metzger, Caldwell \& Schechter & 1992 & AJ, 103, 529  &\\
1&Metzger, Caldwell \& Schechter & 1998 & AJ, 115, 635 & 191\\
 \\[-1pt]
 &Gorynya et al.  & 1992 & Sov. Astron. Lett., 18, 316 & \\
 &Gorynya et al.  & 1998 & Astron. Lett., 24, 815 & \\
 &Gorynya et al.  & 2002 & VizieR On-line Data Catalog: III/229 &\\
2&Gorynya \& Rastorguev   & 2014 & in preparation & 86\\
 \\[-1pt]
 &Pont, Mayor \& Burki  & 1994 & A\&A, 285, 415 & \\
3&Pont et al.  & 1997 &  A\&A, 318, 416& 36\\
 \\[-1pt]
4&Barbier-Brossat, Petit \& Figon  & 1994 & A\&AS, 108, 603 & 7\\
 \\[-1pt]
5 &Malaroda, Levato \& Gallianiet  & 2006 & VizieR On-line Data
 Catalog: III/249 & 2\\
 \\[-1pt]
6&Fernie et al.  & 1995 & IBVS, 4148, 1 & 6\\
 \hline
\end{tabular}
\end{table*}

\section{Morphological evidence}

\subsection{Distribution in the Galactic plane}

The distribution of Cepheids in the Galactic plane can reveal
regions of high gas density which can be associated with spiral
arms or Galactic rings.  Figure~\ref{distrib} shows the
distributions of classical Cepheids and model particles. We can
notice two specific features  here. First, the distribution of
Cepheids is reminiscent of "a tuning fork":  at longitudes
$l>180^\circ$ there is only one spiral arm (the Carina arm)
toward which Cepheids concentrate strongly while at longitudes
$l<180^\circ$ there are two regions with high surface density
located near the Perseus and Sagittarius regions. Second, the
concentration of Cepheids drops sharply in the direction to the
Galactic center while it decreases more gradually in other
directions. These features will be studied further.

\begin{figure} \centering
\resizebox{\hsize}{!}{\includegraphics{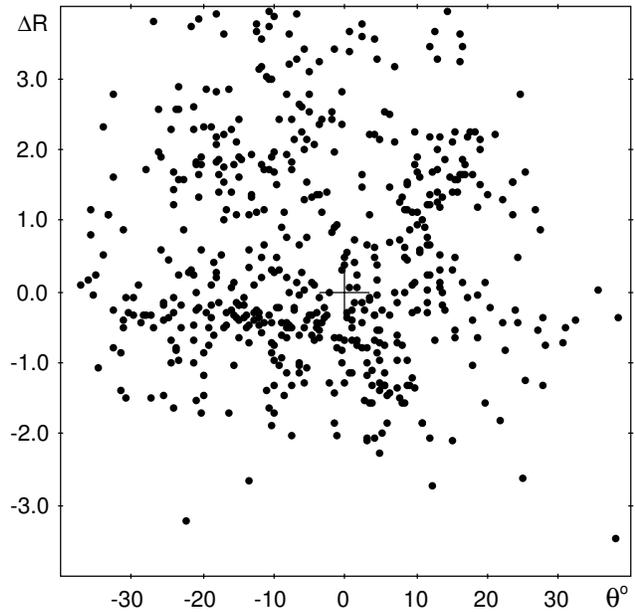}} \caption{
Fork-like structure in the distribution of classical Cepheids
(black circles). Positions of Cepheids are represented in
coordinates ($\theta$, $\Delta R$), where $\theta$ is the
Galactocentric angle and $\Delta R=R-R_0$ is  the difference
between the Galactocentric distances of a Cepheid and the Sun.
The position of the Sun is shown by a cross. We can see the
concentration of Cepheids to the Carina arm at the negative
angles $\theta$ and their concentration to the Perseus and
Sagittarius regions at the positive angles $\theta$.}
\label{fork}
\end{figure}

We also provide Figure~\ref{fork} to emphasize the fork-like
structure in the distribution of Cepheids.  Positions of Cepheids
are represented in coordinates ($\theta$, $\Delta R$), where
$\theta$ is the Galactocentric angle and $\Delta R=R-R_0$ is  the
difference between the Galactocentric distances of a Cepheid and
the Sun.  Cepheids can be seen to  concentrate to the Carina arm
at the negative angles $\theta$ and to the Perseus and
Sagittarius regions at the positive $\theta$.

The position of two outer rings  in the distribution of model
particles can be approximated by two ellipses oriented
perpendicular to each other. The outer ring $R_1$ can be
represented by the ellipse with the semi-axes $a_1=6.3$ and
$b_1=5.8$~kpc, while the outer ring  $R_2$ fits well the ellipse
with $a_2=8.5$ and $b_2=7.6$~kpc. These values correspond to the
solar Galactocentric distance $R_0=7.5$ kpc. The ring $R_1$ is
stretched perpendicular to the bar and the ring $R_2$ is aligned
with the bar, hence the position of the Sun and that of the
sample of Cepheids with respect to the rings is determined by the
Galactocentric angle $\theta_b$ of the Sun  with respect to the
major axis of the bar. We now assume  that Cepheids concentrate
to the outer rings to find the optimum angle $\theta_b$ providing
the best agreement between the position of the rings and the
distribution of Cepheids.

\begin{figure}
\resizebox{\hsize}{!}{\includegraphics{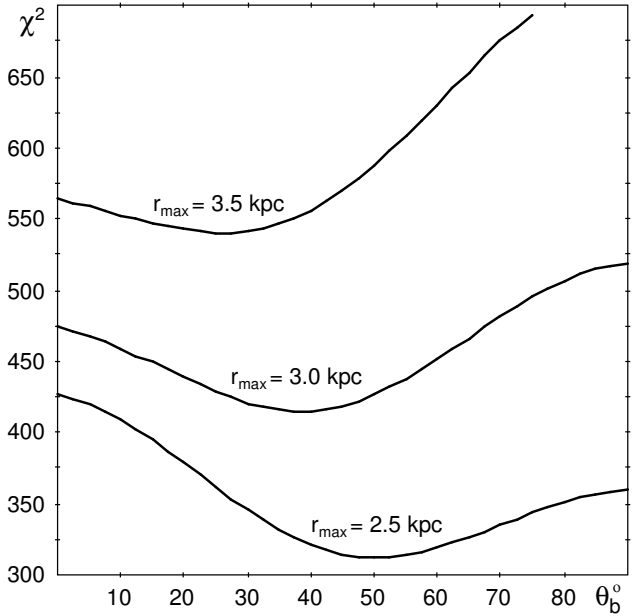}} \caption{The
$\chi^2$ functions calculated for different values of the solar
position angle $\theta_b$ with respect  to the major axis of the
bar.  Cepheids of all periods are included. Three
distance-limited samples of Cepheids confined by $r_{max}=2.5$,
3.0, and 3.5 kpc are considered. The $\chi^2$ functions reach
their minima at $\theta_b=50^\circ$, $37^\circ$, and $25^\circ$,
respectively.} \label{chi2}
\end{figure}

Figure~\ref{chi2} shows three  $\chi^2$ functions -- the sum of
normalized squared deviations of Cepheids from the outer rings
(Press et al.~1987) -- calculated for different values of the
angle $\theta_b$. For each star we determined the minimum
distances to the two ellipses and then took the smallest of the
two values. However, the $\chi^2$ function appears to be very
sensitive to the completeness of the sample. This problem will be
studied further. Here we show the results obtained for three
distance-limited Cepheid samples including stars located within
$r_{max}=2.5$, 3.0, and 3.5 kpc from the Sun. Table 3 lists the
parameters of different samples: the number $N$ of Cepheids, the
minimal value $\chi^2_{min}$, the standard deviation $\sigma_0$
of a Cepheid from the model distribution, and the angle
$\theta_b$ corresponding to $\chi^2_{min}$.  In this section we
consider the parameters derived without any segregation over
periods, deferring the discussion of the values obtained   for
short- and long-period Cepheids separately to section 3.2. The
first three rows of Table~3 indicate that the $\chi^2$ functions
reach their minima at $\theta_b=50^\circ$, 37$^\circ$, and
25$^\circ$, respectively. The random errors of these estimates
are of about $\pm5^\circ$. Note that the different values of
$\chi^2_{min}$ represented in Table 3 are derived for different
samples and cannot be compared with each other. This also holds
true for the values of $\sigma_0$.

The  position angle estimates $\theta_b=50\pm5^\circ$ and
$37\pm5^\circ$ derived for the samples within $r_{max}=2.5$ and
3.0 kpc, respectively, agree well with the estimate
$\theta_b=45\pm5^\circ$ obtained from the kinematics of
OB-associations (Mel'nik \& Rautiainen 2009; Rautiainen \&
Mel'nik 2010). It must be  "the fork-shaped" distribution of
Cepheids that determines the angle $\theta_b$ being close to
45$^\circ$. At  longitudes $l>180^\circ$ two outer rings fuse
together to form one spiral fragment -- the Carina arm, whereas
at longitudes $l<180^\circ$ two outer rings are prominent,
generating the fragments of the Sagittarius and Perseus arms.

We can see  that  the estimates of $\theta_b$ decrease with
increasing $r_{max}$ (Table 3). This shift  can be attributed to
numerous stars scattered in the direction of the anti-center at
distances $r>2$ kpc (Figures~\ref{distrib},\ref{fork}). Their
inclusion into the sample makes the outer ring $R_2$ to be
aligned with the line connecting the Sun and the Galactic center.
As the ring $R_2$ is aligned with the bar, the increase of
$r_{max}$ must be accompanied by the decrease in $\theta_b$.

On the whole, the optimum position angle  $\theta_b$ providing
the best agreement between the distribution of Cepheids  inside
$3\pm0.5$ kpc and the model of the outer ring $R_1R_2'$ can be
derived by averaging the  $\theta_b$ estimates listed in  the
first three rows of Table 3, yielding $\theta_b=37^\circ$. The
average scatter of these estimates  is $\pm 12^\circ$. Assuming
that two errors in the determination of $\theta_b$, $\pm12$ and
$\pm5$, are independent, we can estimate the combined error as
$\varepsilon^2=\varepsilon_1^2+\varepsilon_2^2$, or $\sim
13^\circ$.

\begin{table}
 \caption{Parameters of different samples of Cepheids}
 \begin{tabular}{lccccc}
  \\[-13pt] \hline\\[-7pt]
 Period &$r_{max}$ & $N$  & $ \chi^2_{min}$ &   $\sigma_0$ & $\theta_b$  \\
  \\[-7pt] \hline\\[-7pt]
 &2.5 kpc & 314 & 312.15 &0.72 kpc & $50\pm5^\circ$\\
 All&3.0 kpc & 372 & 415.17 &0.81 kpc & $37\pm5^\circ$\\
periods&3.5 kpc & 413 & 540.57 &0.90 kpc & $25\pm5^\circ$\\
  \\[-7pt] \hline\\[-7pt]
$P<8$ d&2.5 kpc & 242 & 235.88 &0.71 kpc & $53\pm5^\circ$\\
$P<8$ d&3.0 kpc & 281 & 358.64 &0.81 kpc & $40\pm5^\circ$\\
$P<8$ d&3.5 kpc & 326 & 521.97 &0.91 kpc & $28\pm5^\circ$\\
  \\[-9pt] \hline\\[-7pt]
 $P>8$ d&2.5 kpc &  72 &  75.33 &0.74 kpc & $43\pm10^\circ$\\
$P>8$ d&3.0 kpc &  91 & 106.80 &0.78 kpc & $30\pm10^\circ$\\
$P>8$ d&3.5 kpc & 105 & 153.16 &0.87 kpc & $23\pm10^\circ$\\
 \hline
\end{tabular}
\end{table}

\newpage

\subsection{Distribution over periods}

There is a wide-spread opinion that  only classical Cepheids with
long periods   are suitable for  study of the Galactic structure.
Note that  Cepheids with the periods $P<4$ days actually require
more thorough study for classification. First, some  of them are
oscillating in the first overtone rather than in the fundamental
tone and this fact should be taken into account when computing
the distances. Second, short-period classical Cepheids can be
confused with W Vir type variables which  are old stars of
population II. However, classical Cepheids have very specific
features in light curves (Hertzsprung Progression) (Hertzsprung,
1926) which distinguish them from Cepheids of other types.

The subdivision of Cepheids into long- and short-period groups is
a matter of convention. Usually, they are separated by the period
of 10 days. However, we believe that the value of 8 days is more
appropriate. Figure~5 shows the distribution of  fundamental
periods of Cepheids. We can see a clear maximum in the
distribution of Cepheids with periods $P<8$ days and almost a
plateau distribution for Cepheids with $P>8$.

Figure~\ref{cepd} shows the  distribution of classical Cepheids
with short ($P<8$ d) and long ($P>8$ d) periods in the Galactic
plane. It shows a conspicuous lack of long-period Cepheids in
quadrant III. However, short- and long-period Cepheids generally
show a similar behavior: they concentrate to the Carina arm in
 quadrant IV, to the Sagittarius region in  quadrant I, and
to the Perseus region in  quadrant II.

We repeat our study of the  $\chi^2$ functions  for short- and
long-period Cepheids separately.  Table~3 presents the parameters
derived for Cepheids with periods $P<8$ and $P>8$ days in three
regions: $r_{max}=2.5$, 3.0, and 3.5 kpc.  We can see that the
angles $\theta_b$ corresponding to $\chi^2_{min}$ derived for the
short- and long-period Cepheids in the same regions coincide
within the errors. On the whole, short- and long-period Cepheids
demonstrate the same tendency.

\begin{figure}
\resizebox{\hsize}{!}{\includegraphics{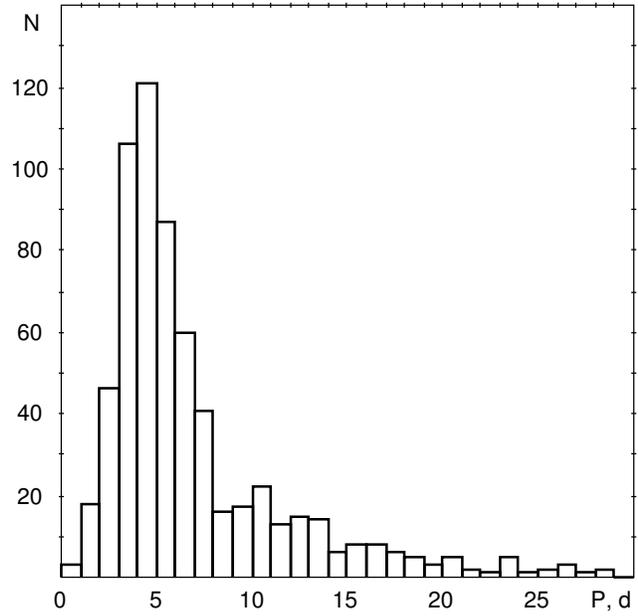}} \caption{The
number distribution of classical Cepheids from the catalog by
Berdnikov et al. (2000) over fundamental periods $P$.}
\label{per}
\end{figure}
\begin{figure}
\resizebox{\hsize}{!}{\includegraphics{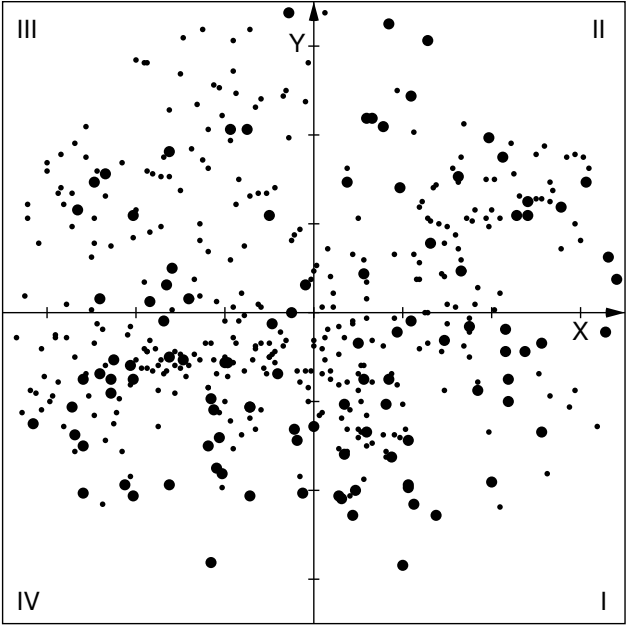}} \caption{ The
distribution of classical Cepheids with short (small circles) and
long (large circles) periods in the Galactic plane within 3.5 kpc
from the Sun. Periods are divided into two groups with respect of
$P=8$ days. Roman numbers indicate quadrants. We can see the
conspicuous lack of long-period Cepheids in  quadrant III.}
\label{cepd}
\end{figure}
\newpage
\begin{figure*} \centering
\resizebox{14 cm}{!}{\includegraphics{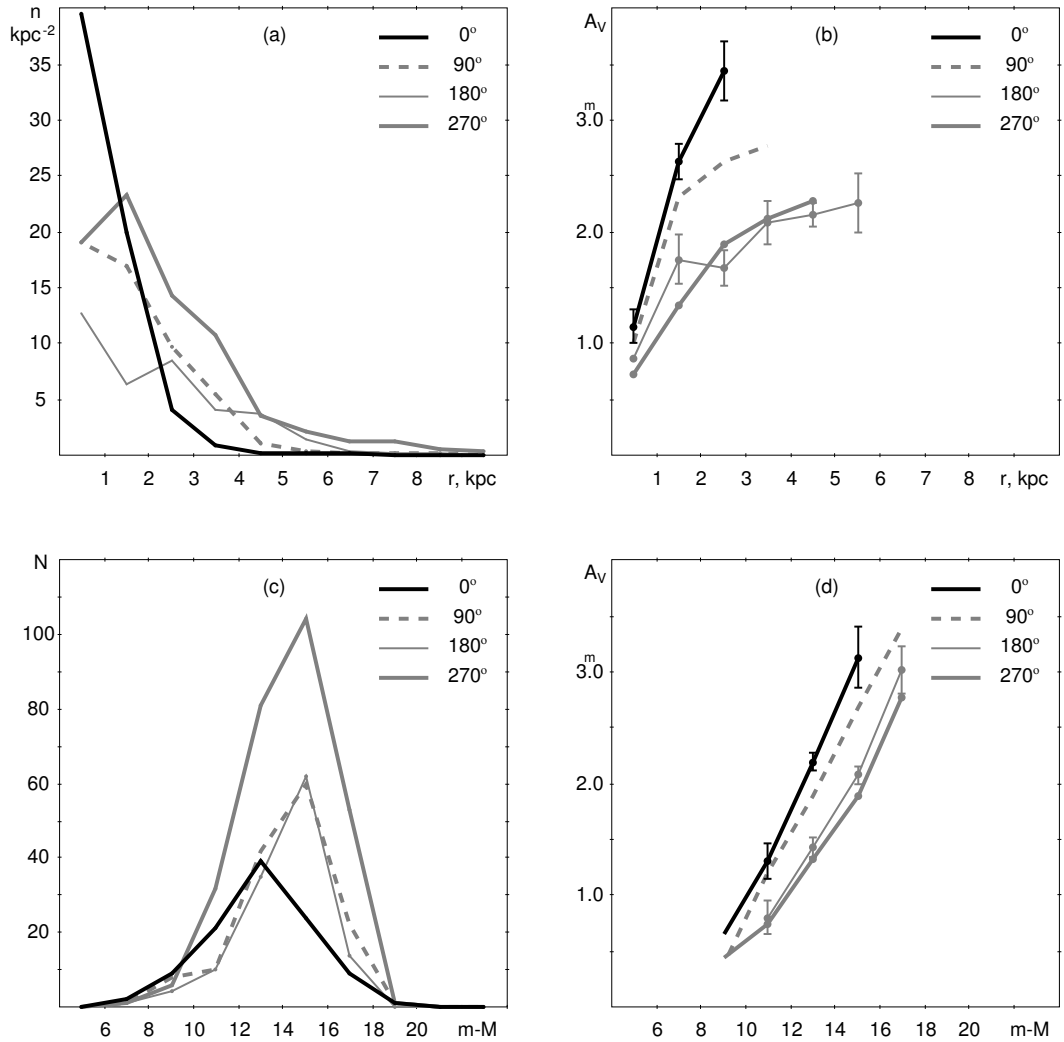}} \caption{(a)
Variations of the average surface density $n$ of Cepheids  along
the heliocentric distance $r$. (b) Variations of  the average
extinction $A_V$ along the  distance $r$. (c) Distribution of
number of Cepheids $N$ by the apparent distance modulus $m-M$,
where $N$ is calculated in 2$^m$-wide intervals of distance
modulus. (d) Growth of  $A_V$ with the increasing $m-M$.
Different lines show the values calculated for four sectors
confined by the longitudes: $l=0\pm45^\circ$, $l=180\pm45^\circ$,
$l=90\pm45^\circ$, and $l=270\pm45^\circ$. Frame (a) demonstrates
the sharp drop of $n(r)$ in the direction of the Galactic center
$|l|<45^\circ$ (black solid line).  Frame (c) shows that the
distribution of $N$ in the direction of the Galactic center peaks
at 13$^m$, whereas those for the other sectors peak at 15$^m$.
The bars in frames (b) and (d) indicate the errors of $A_V$ which
are shown only for the direction of the Galactic center and
anti-center, in other sectors they are less than 0.2$^m$. It is
evident from frames  (b) and (d)  that  extinction $A_V$ in the
direction of Galactic center $|l|<45^\circ$  is greater than in
other sectors.} \label{extin_all}
\end{figure*}

\subsection{Average surface density and extinction in different directions}

The average surface density  and interstellar extinction of
Cepheids in different directions  can provide  us with additional
information about the distribution of gas and dust in the Galaxy.
To study these characteristics, we  subdivide the sample of
Cepheids into four sectors: the directions to the center
$|l|<45^\circ$ and anti-center $|l-180|<45^\circ$, those in the
sense of Galactic rotation $|l-90|<45^\circ$ and in the opposite
sense $|l-270|<45^\circ$. For each sector we calculate the number
of Cepheids in annuli of width $\Delta r=1$ kpc. The ratio of the
number of Cepheids in a quarter of an annulus to its area gives
us the average  surface density $n$ of Cepheids per kpc$^2$. This
value depends on the sector and heliocentric distance $r$. We
also calculate the average $V$-band extinction $A_V$ for Cepheids
in different annuli and in different sectors.  We compute the
$A_V$ value for each star as $A_V=3.26 E_{B-V}$ (Berdnikov et al.
1996a), where $E_{B-V}$ is the color excess from the catalog by
Berdnikov et al.~(2000). Note that the values of $A_V$ were not
used in the determination of Cepheid distances which were derived
from the $K$-band magnitudes and extinction $A_K$ (see also
section 2).

Figure~\ref{extin_all} (a) and (b) shows the variations of $n(r)$
and $A_V(r)$ along the heliocentric distance $r$ calculated for
the four sectors. The surface density of Cepheids $n$ in the
direction of the center ($|l|<45^\circ$) can be seen to drop
sharply with increasing $r$: it decreases by a factor of $\sim10$
from the distance range of 0--1 kpc to that of 2--3 kpc. For
comparison, this ratio amounts only 1.3--1.9 for the other three
sectors. The surface density of Cepheids in the direction of the
anti-center $|l-180|<45^\circ$ is less than in other directions
within 4 kpc, for larger distances the estimates of $n$ are very
uncertain. Extinction $A_V$ can also be seen to increase sharply
toward the Galactic center ($|l|<45^\circ$).

Additionally, we consider the variations in  the number of
Cepheids $N$ and $A_V$ along the apparent distance modulus $m-M$
(Figure~\ref{extin_all}cd), where $N$ and the average extinction
$A_V$ are calculated in 2$^m$-wide  distance modulus bins. The
distribution of $N$ for the sector $|l|<45^\circ$ peaks at
13$^m$, whereas those for the other sectors peak   at 15$^m$. It
means that Cepheids can generally be discovered out to $m-M=15^m$
with the current instruments. Probably,  the number of Cepheids
$N$ in the sector $|l|<45^\circ$  drops considerably beyond
13$^m$  due to their lower surface density there. However, the
alternative explanation is that crowding is much stronger in the
direction of the Galactic center $|l|<45^\circ$, preventing the
discovery of Cepheids.  Possibly, the survey VISTA variables in
the Via Lactea  (Saito et al. 2012) would help solve this
alternative.

The bars in  Figure~\ref{extin_all} (b) and (d) indicate the
errors in extinction $A_V$ which are shown only for the direction
of the Galactic center $|l|<45^\circ$  and anti-center
$|l-180|<45^\circ$. In other sectors, the errors of $A_V$  are
less than 0.2$^m$. We can see that extinction $A_V$ in the
direction of the Galactic center $|l|<45^\circ$ is greater than
in other directions. Particularly,  for distance moduli
$m-M>11^m$ ($r>1.5$ kpc), the  extinction $A_V$ in the direction
$|l|<45^\circ$ is greater than in the opposite direction
$|l-180|<45^\circ$ at a significance level of $P>2\sigma$.

Generally, these particularities can be explained in terms of the
model of the Galaxy with the outer rings located near the Sun. It
is probable that the line of sight directed toward  the Galactic
center intersects the outer ring $R_1$ (Figure~\ref{distrib})
which may contain a great amount of dust. Note that N-body
simulations show that the ring $R_1$ forms not only in the gas
subsystem but also in the stellar component (Rautiainen \& Salo
2000). We can expect that a high concentration of  old stars in
the ring $R_1$ can be accompanied by  a great  amounts of dust
there.

\subsection{Completeness of the sample}

To estimate the radius of completeness of our sample, we need a
hypothesis about the physical distribution of Cepheids. On the
one hand, Cepheids  should not be distributed uniformly in the
Galactic disk, they must concentrate to  spiral arms or to
Galactic rings. On the other hand, Cepheids at larger distances
are less likely to be discovered than those situated close to the
Sun.  Let us suppose that  spiral arms or outer rings can
increase the surface density of Cepheids not more than by a
factor of two. That gives us a simple criterion: the sample is
complete until the surface density n(r) drops not more than to
half its maximum value. When applied to the four sectors this
criterion suggests that our sample is complete out to 2 kpc in
the direction of the Galactic center ($|l|<45^\circ$), out to  3
kpc in two directions: that of the anti-center $|l-180|<45^\circ$
and in the sense of Galactic rotation $|l-90|<45^\circ$, and out
to $\sim4$ kpc in the sense opposite that of Galactic rotation
$|l-270|<45^\circ$ (Figures~\ref{extin_all}a). The latter result
may be due to the great interest of observers in the Carina arm
and its extension.

Another criterion of the completeness of the sample can be
formulated on the basis of variations in the number of Cepheids
$N$ along the visual distance modulus $m-M$
(Figure~\ref{extin_all}b). Suppose that the sample is complete
until the number of Cepheid $N$ increases with increasing $m-M$.
This establishes the limits of $m-M=13^m$ for the sector in the
direction of the Galactic center ($|l|<45^\circ$) and $m-M=15^m$
for the three other sectors. To transform these apparent distance
moduli into distances, we must make an assumption about
extinction. The variation of $A_V$ with distance modulus
(Figure~\ref{extin_all}d) shows that the average extinction at
$m-M=13^m$ in the  sector ($|l|<45^\circ$) is nearly 2.0$^m$ and
it is also close to $A_V=2.0^m$ at $m-M=15^m$ in three other
sectors. So we can suppose that a Cepheid will be discovered if
its true distance modulus is $(m-M)_0=11.0^m$ in the sector
($|l|<45^\circ$) and $(m-M)_0=13.0^m$ in other sectors, i.e if it
is located within $r=1.6$ kpc and $r=4.0$ kpc, respectively which
is consistent with our previous estimates.

Let us consider  another  criterion of completeness which is
based on the  distribution  of apparent magnitudes (Szabados
2003, for example). Figure~\ref{mv} shows the distribution of
apparent visual magnitudes $m_v$ of Cepheids from the catalog by
Berdnikov et al. (2000). The number of Cepheids can be seen to
increase with  $m_v$ till the value 12$^m$. Note that the average
apparent magnitudes $m_v$ of Cepheids in the heliocentric
distance intervals 2--3 and 3--4 kpc are 11 and 12$^m$,
respectively. So again we can suppose that the catalog by
Berdnikov et al. (2000) is  nearly complete till $\sim 3$ kpc. To
formulate more precisely, the incompleteness is getting obvious
at $r>3$ kpc.

\section{Kinematical evidence}

\subsection{Rotation curve}

 \begin{table*}
 \centering
 \caption{\bf Parameters of the rotation curve and the
solar motion}
 \begin{tabular}{lcccccccl}
  \\[-7pt] \hline\\[-7pt]
  Objects &$\Omega_0$ & $\Omega'_0$ & $\Omega''_0$ & $u_0$ & $v_0$ &  A & $\sigma_0$ & N \\
  & km s$^{-1}$  & km s$^{-1}$  & km s$^{-1}$  &
  km s$^{-1}$ & km s$^{-1}$  &km s$^{-1}$ &km s$^{-1}$ &   \\
  &   kpc$^{-1}$ & kpc$^{-2}$ & kpc$^{-3}$ &
   & & kpc$^{-1}$& &  \\
  \\[-7pt] \hline\\[-7pt]
 Cepheids & 28.8 & -4.88 & 1.07 & 8.1 & 12.7 &  18.3 & 10.84 & 474\\
 all periods& $\pm0.8$ & $\pm0.14$ & $\pm0.17$ & $\pm0.8$ & $\pm1.0$  & $\pm0.6$ &\\
  \\[-7pt] \hline\\[-7pt]
 Cepheids & 30.0 & -4.90 & 0.92 & 8.3 & 12.1 &  18.4 & 11.19 & 323\\
 $P<8$d& $\pm1.0$ & $\pm0.22$ & $\pm0.25$ & $\pm1.0$ & $\pm1.2$  & $\pm0.8$ &\\
  \\[-7pt] \hline\\[-7pt]
 Cepheids & 27.0 & -4.85 & 1.31 & 7.7 & 14.2 &  18.2 & 10.04 & 151\\
 $P>8$d& $\pm1.1$ & $\pm0.19$ & $\pm0.24$ & $\pm1.3$ & $\pm1.7$  & $\pm0.7$ &\\
  \\[-7pt] \hline\\[-7pt]
 OB-associations & 30.6 & -4.73 & 1.43 & 7.7 & 11.6 &  17.7 & 7.16 & 132\\
 & $\pm0.9$ & $\pm0.18$ & $\pm0.21$ & $\pm1.0$ & $\pm1.3$  & $\pm0.7$ &\\
 \hline
\end{tabular}
\end{table*}

Before studying the non-circular motions, we must determine the
main parameters of the circular rotation of our objects. Using the rotation
curve derived from the same sample of objects allows us to avoid
systematical effects due to the eventual inconsistency of the distance scales of
two different samples.

We determine the parameters of the rotation curve based only on
the Cepheids located within 3.5 kpc from the Sun ($r<3.5$ kpc)
and within 0.5 kpc ($|z|<0.5$ kpc) from the Galactic plane. This
subsample includes 257 Cepheids with available accurate
line-of-sight velocities and 217 Cepheids with Hipparcos proper
motions. The approach we use to solve the Bottlinger equations
was described in detail in our earlier papers (Dambis et
al.~1995; Mel'nik, Dambis \& Rastorguev 1999; Mel'nik \& Dambis
2009), so we do not repeat it here. We list the inferred
parameters of the rotation curve and solar motion in Table~4,
where $\Omega_0$ is the angular rotation velocity $\Omega(R)$ at
$R=R_0$; $\Omega'_0$ and $\Omega''_0$ are its first and second
derivatives taken at $R=R_0$; $u_0$ and $v_0$ are the components
of the solar motion with respect to the centroid of  the sample
in the direction toward the Galactic center and Galactic
rotation, respectively; $N$ is the number of conditional
equations.  Table~4 also includes the parameters derived for
short- and long-period Cepheids separately and those for
OB-associations.

\begin{figure}
\resizebox{\hsize}{!}{\includegraphics{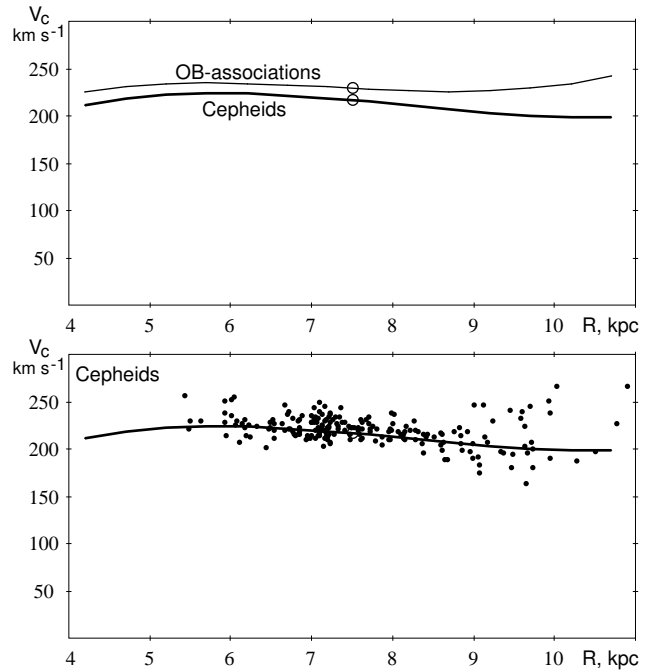}} \caption{Top
panel: the Galactic rotation curve derived from an analysis of
line-of-sight velocities and proper motions of Cepheids (thick
line) and those of  OB-associations (thin line). The position of
the Sun is shown by a circle. Lower panel: the scatter of
individual azimuthal velocities of Cepheids  with respect to the
rotation curve. It is built for Cepheids of all periods.}
\label{rot_curve}
\end{figure}

\begin{figure}
\centering \resizebox{8 cm}{!}{\includegraphics{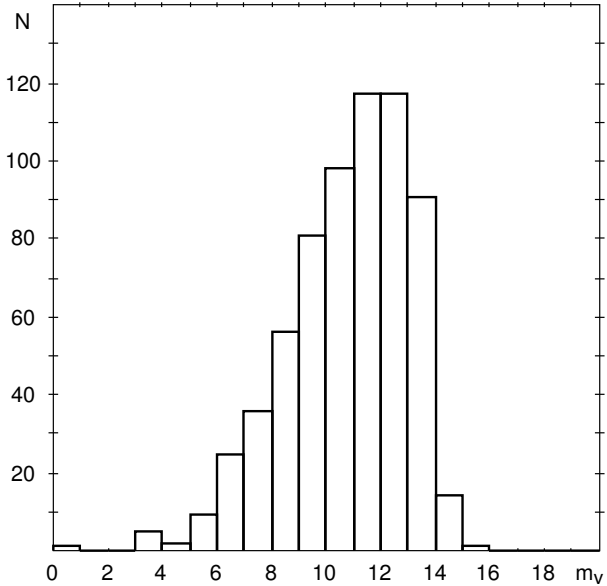}}
\caption{The  distribution apparent magnitudes $m_v$ of classical
Cepheids.} \label{mv}
\end{figure}

Figure~\ref{rot_curve} (top panel) shows the rotation curve
derived from an analysis of the line-of-sight velocities and
proper motions of Cepheids  of all periods. For comparison, we
also show the rotation curve based on the data for
OB-associations (Mel'nik \& Dambis 2009). The parameters of the
rotation curve derived for Cepheids and OB-associations agree
within their errors,   see Table~4. Our Cepheid sample yields
$\Omega_0=28.8\pm0.8$ km s$^{-1}$ kpc$^{-1}$ which  is also
consistent with  other studies of the Cepheid kinematics:
$\Omega_0=27.2\pm0.9$ (Feast \& Whitelock 1997) and
$\Omega_0=27.5\pm0.5$ (Bobylev \& Baikova 2012). However, there
is a systematical difference between the value of $\Omega_0$
obtained for  Cepheids and that inferred for OB-associations and
maser sources, $\Omega_0=31\pm1$ km s$^{-1}$ (Reid et al.~2009a,
Mel'nik \& Dambis 2009; Bobylev \& Baikova 2010). The linear
rotation velocity at the solar Galactocentric distance estimated
from the kinematics of Cepheids is systematically lower than the
value derived from OB-associations. Moreover, the rotation curve
of Cepheids seems to be slightly descending, whereas that of
OB-associations is nearly flat within the 3 kpc neighborhood of
the Sun. We are inclined to attribute these differences to the
fact that the distances and proper motions available for Cepheids
are less accurate than those of OB-associations. Averaging the
distances and proper motions of stars within each OB-association
may have given a considerable advantage. This problem requires
further study.

The lower panel of Figure~\ref{rot_curve} shows the scatter of
individual azimuthal velocities  of Cepheids with respect to the
rotation curve.  The standard deviation  of the projected
velocities onto the Galactic plane from the rotation curve is
$\sigma_0=10.8$ km s$^{-1}$ for Cepheids of all periods.

We also calculated the parameters of the rotation curve for
short- and long-period Cepheids (Table~4). It can be seen  that
parameters determined for  Cepheids with $P<8$ and $P>8$ days are
consistent within  the errors except for $\Omega_0$, which is
conspicuously smaller for long-period Cepheids. Note that the
standard deviation of the velocities from the rotation curve
equals $\sigma_0=11.19$ and 10.04 km s$^{-1}$ for short- and
long-period Cepheids, respectively. The small difference between
them suggests that both groups of Cepheids are suitable for study
the Galactic structure.


\newpage
\begin{figure*} \centering
\resizebox{\hsize}{!}{\includegraphics{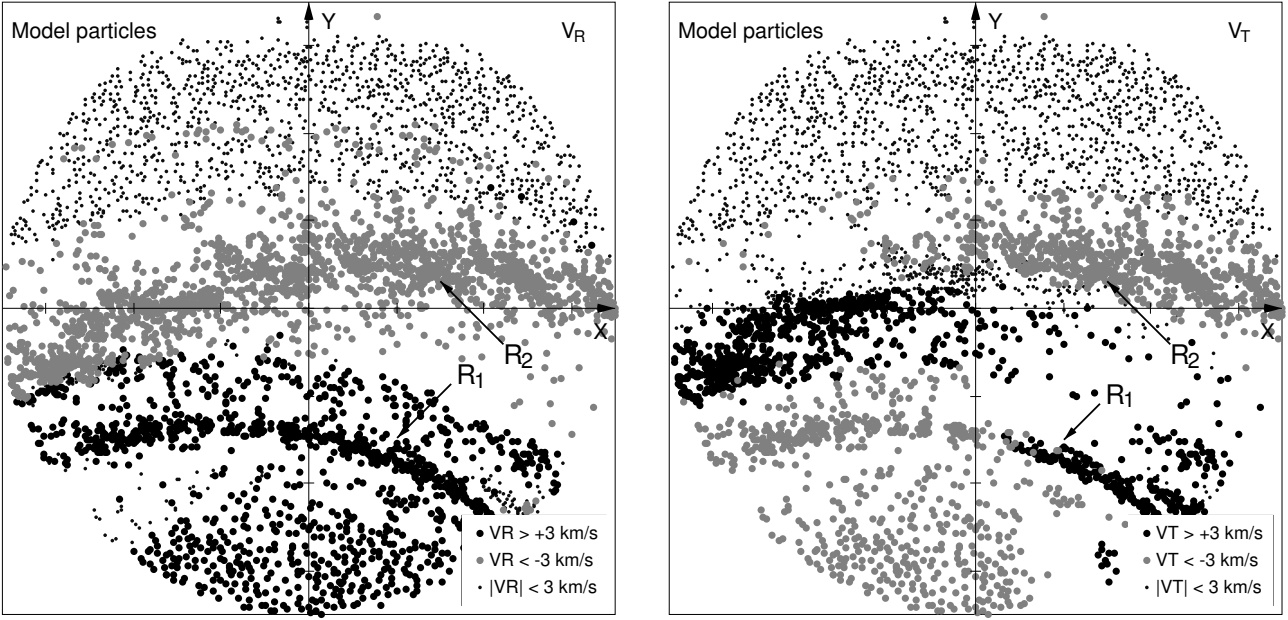}}
\caption{Distribution of negative and positive residual
velocities $V_R$ (left panel) and $V_T$ (right panel) calculated
for  model particles located in the solar neighborhood of 3.5
kpc.  The radial velocities $V_R$ are subdivided into three
groups: negative ($V_R<-3$ km s$^{-1}$), positive ($V_R>+3$ km
s$^{-1}$) and close to zero ($|V_R|<3$ km s$^{-1}$). The same was
done for the azimuthal velocities $V_T$. The solar position angle
$\theta_b$ for model particles is chosen to be
$\theta_b=45^\circ$. The arrows show the locations of the
fragments of the outer rings $R_1$ and $R_2$: the ring $R_1$ is
closer to the center than the ring $R_2$. The $X$-axis is
directed in the sense of the Galactic rotation, the $Y$-axis is
directed away from the Galactic center. The Sun is in the origin.
One tick interval along the $X$- and $Y$-axis corresponds to 1
kpc. } \label{VR_VT_gray}
\end{figure*}

\begin{figure*}
\resizebox{\hsize}{!}{\includegraphics{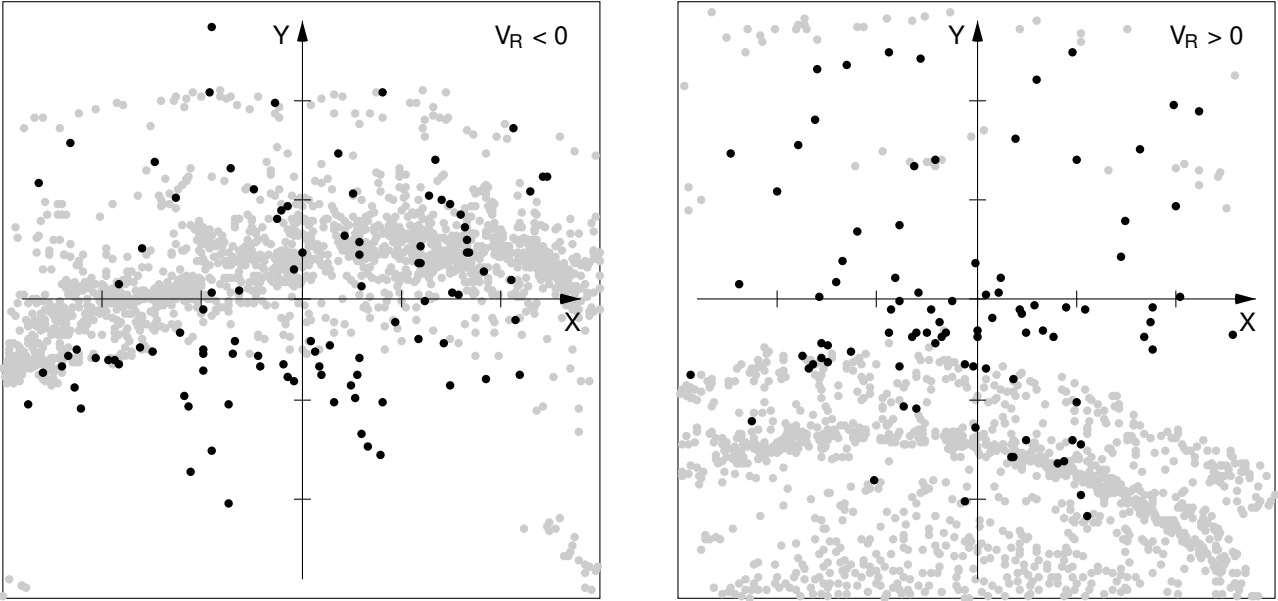}}
\caption{Distribution of  Cepheids (black circles) and model
particles (gray circles) with negative ($V_R<0$) and positive
($V_R>0$) radial residual velocities in the Galactic plane. Left:
objects  with $V_R<0$ are supposed to belong to the descending
segment of the outer ring $R_2$. Right:  objects with the
positive radial residual velocities ($V_R>0$) are mostly located
in  quadrants I and IV. The $X$-axis is directed in the sense of
the Galactic rotation, the $Y$-axis is directed away from the
Galactic center. The Sun is at the origin. One tick interval
along the $X$- and $Y$-axis corresponds to 1 kpc.}
\label{VR_neg_pos}
\end{figure*}

\subsection{Residual velocities of Cepheids and model particles}

Residual velocities characterize  non-circular motions in the
Galactic disk. We calculate the residual velocities for Cepheids
as the differences between the observed heliocentric velocities
and the computed velocities due to the circular rotation law and
the adopted components of the solar motion defined by the
parameters listed in  Table~4 (first row). For model particles
the residual velocities are determined  with respect to the model
rotation curve. We consider the residual velocities in the radial
$V_R$ and azimuthal $V_T$ directions. Positive radial residual
velocities $V_R$ are directed away from the Galactic center while
positive azimuthal residual velocities $V_T$ are in the sense of
Galactic rotation.

Figure~\ref{VR_VT_gray} shows the distribution of  model
particles with  negative and positive residual velocities $V_R$
and $V_T$ located within  3.5 kpc from the Sun. The left panel
demonstrates the distribution of radial residual velocities
$V_R$. Model particles  in the outer ring $R_1$ (the one which is
closer to the Galactic center) have positive velocities $V_R$
(black circles), whereas those in the ring $R_2$ have negative
velocities $V_R$ (gray circles). Particles located at the
Galactic periphery ($R>R_0+2$ kpc) have close to zero velocities
$V_R$ (black points). The right panel shows the distribution of
azimuthal velocity $V_T$. Model particles  in the ring $R_1$
lying  left of the Sagittarius complex ($l\approx 15^\circ$,
$r\approx 1.5$ kpc) have mostly negative $V_T$ velocities while
those situated  right of it have mostly positive $V_T$
velocities. It is not a chance coincidence because the position
angle of the bar $\theta_b$ was chosen in such a way that model
particles reproduce nearly zero velocities $V_T$ of
OB-associations in the Sagittarius complex (Mel'nik \& Rautiainen
2009). In the ring $R_2$ we see the opposite velocity gradient:
model particles located  at the negative x-coordinates have
mostly positive $V_T$ velocities (black circles), whereas those
situated at the positive x-coordinates have mainly negative $V_T$
velocities (gray circles).

Let us consider the distribution of Cepheids and model particles
with negative ($V_R<0$) and positive ($V_R>0$) radial residual
velocities (Figure~\ref{VR_neg_pos}). The left panel shows the
distribution of objects with negative radial velocities $V_R$
which are supposed to belong to the outer ring $R_2$. Within
$r<3.0$ kpc from the Sun, the elliptic ring $R_2$ can be
represented as a fragment of the spiral arm. Its   pitch angle
appears to be $i=8.3\pm3.9^\circ$ and $i=6.0\pm0.5^\circ$ for
Cepheids and model particles, respectively. The positive value of
the pitch angle $i$ indicates that the spiral arm is leading and
corresponds to the solar position near the descending segment of
the outer ring $R_2$. Since the    ring $R_2$ is aligned with the
bar, the location of the Sun near the descending segment of the
outer ring $R_2$ reflects the well-known fact that the bar's end
closest to the Sun lies in  quadrant I.

Figure~\ref{VR_neg_pos} (right panel) shows Cepheids and model
particles  with positive  radial residual velocities ($V_R>0$).
These objects are expected to concentrate to the ascending
segment of the outer ring $R_1$. Unfortunately, we  see no good
agreement here. However, Cepheids do not occupy all the 3-kpc
solar neighborhood, they as well as  model particles with $V_R>0$
are mostly located at negative $y$-coordinates.

\begin{figure}
\resizebox{\hsize}{!}{\includegraphics{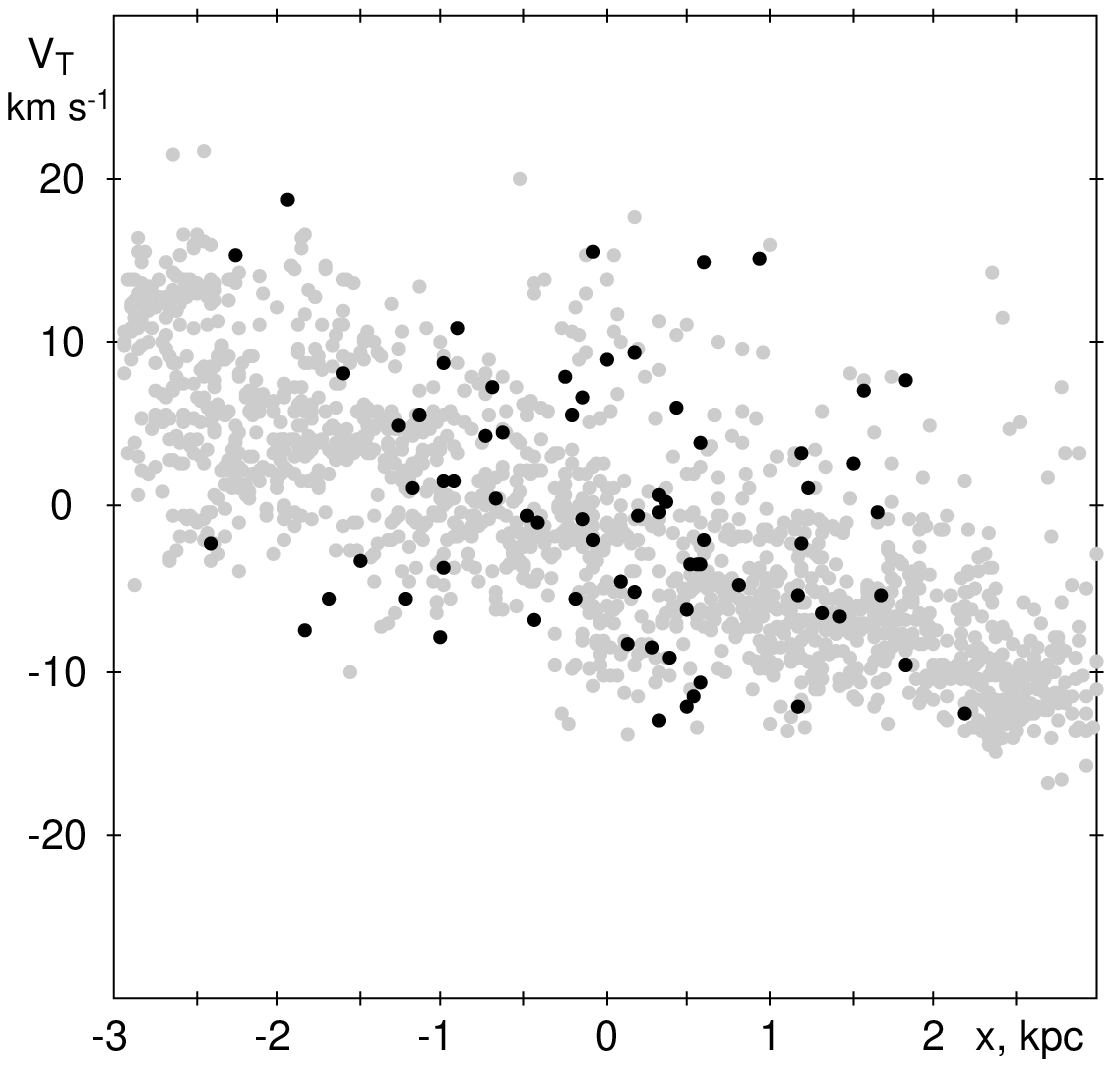}}
\caption{Dependence of azimuthal velocity $V_T$ on coordinate $x$
for the Cepheids (black circles) and model particles (gray
circles) with negative radial residual velocities ($V_R<0$).
These objects are supposed to belong to the descending segment of
the outer ring $R_2$. Both Cepheids and model particles
demonstrate a decrease in $V_T$  with increasing $x$.}
\label{VT-x}
\end{figure}

Figure~\ref{VT-x} shows the dependence of  azimuthal velocity
$V_T$ on  coordinate $x$  for Cepheids and model particles with
negative radial velocities ($V_R<0$). The objects studied are
located within 3 kpc of the Sun and are supposed to belong to the
descending segment of the outer ring $R_2$.  We consider only
Cepheids with  small errors ($\varepsilon_{vl}<10$ km s$^{-1}$)
of the tangential velocity  in the Galactic plane $V_l$. The
value of 10 km s$^{-1}$ corresponds to the average deviation of
the Cepheid velocity from rotation curve ($\sigma_0$ in Table 4).
The error $\varepsilon_{vl}$ is determined by the error of the
proper motion in the Galactic plane $\mu_l$ and the distance $r$,
$\varepsilon_{vl}=4.74r\varepsilon_{\mu_l}$, where proper motion
is in mas yr$^{-1}$ and the distance is in kpc. We use the linear
law to describe the $V_T$-$x$ relation: $V_T=ax+b$. The slope $a$
for Cepheids and model particles appears to be $a=-2.0\pm0.8$ and
$a=-3.3\pm0.1$, respectively. Both values of $a$ are negative,
and this trend for Cepheids has a significance level of
$P>2\sigma$.

Generally, the kinematical features obtained for Cepheids with
negative radial velocities $V_R$ are consistent with the solar
location near the descending segment of the outer ring $R_2$.

\section{Discussion and conclusions}

We use the data from the catalog by Berdnikov et al.~(2000) to
study the distribution and kinematics of classical Cepheids  in
terms of the model of the Galactic ring $R_1R_2'$ (Mel'nik \&
Rautiainen 2009). The best agreement between the distribution of
Cepheids   located within $r_{max}=2.5$, 3.0, and 3.5 kpc from
the Sun and the model of the outer  ring $R_1R_2'$ is obtained if
the position angle of the Sun with respect to the bar major axis
is $\theta_b=50\pm5^\circ$, $37\pm5^\circ$, and $25\pm5^\circ$,
respectively. Averaging these values gives the final estimate of
$\theta_b=37\pm13^\circ$.  It is "the fork-like" structure in the
distribution of Cepheids that determines the angle $\theta_b$
being close to 45$^\circ$: at longitudes $l>180^\circ$ two outer
rings fuse together to form one spiral fragment -- the Carina
arm, whereas at longitudes $l<180^\circ$ two outer rings run
separately producing  the Sagittarius and Perseus arm-fragments.

To study the surface density $n$ and extinction $A_V$ of Cepheids
in different directions, we subdivide the sample into four
sectors and calculate the average values of $n$ and  $A_V$ at
different heliocentric distances $r$. The surface density $n$ of
Cepheids  appears to drop sharply in the direction of the
Galactic center ($|l|<45^\circ$). Furthermore, extinction $A_V$
grows most rapidly just in this direction. These features can be
due to the  presence of the   ring $R_1$ located in the direction
of the Galactic center  at 1--2 kpc from the Sun.

Our analysis of variations in the surface density of Cepheids
along distance  gives us the estimate of the completeness radius
of the sample which appears to lie in the 2-4 kpc interval for
the different sectors.  A similar result was obtained from an
analysis of distance moduli  and visual magnitudes. We adopt
$3\pm1$ kpc as the average value. Probably, within 3 kpc from the
Sun, the apparent distribution of known Cepheids reflects their
physical distribution rather than instrumental-related
difficulties in their discovery and study.

The parameters of the rotation curve derived from the samples of
Cepheids  and OB-associations (Mel'nik \& Dambis 2009) are
consistent within the errors (Table~4).

We study the distribution of  Cepheids and model particles  with
negative radial residual velocities ($V_R<0$), which  inside 3
kpc of the Sun must belong to the outer   ring $R_2$. The
selected Cepheids and model particles demonstrate  similar
distribution in the Galactic plane: both samples concentrate to
the fragment of the leading spiral arm  with the pitch angle of
$i=8.3\pm3.9^\circ$ and $i=6.0\pm0.5^\circ$, respectively. A
similar leading fragment was found in the distribution of
OB-associations with negative radial residual velocities
($V_R<0$) (Mel'nik 2005). The appearance of the leading fragment
suggests that the Sun is located near the descending segment of
the   ring $R_2$. Moreover, selected Cepheids and model particles
exhibit similar variations of azimuthal velocity $V_T$ in the
direction of Galactic rotation (the  $x$ coordinate).

All this morphological and kinematical evidence suggests the
existence of a  ring $R_1R_2'$ in the Galaxy. N-body simulations
show that the descending segments of the outer rings $R_2$ often
include clumps and spurs  (Rautiainen \& Salo 2000). Probably,
the Galactic outer   ring $R_2$ is not homogeneous in the solar
neighborhood. The local Cygnus arm ($l=70\textrm{--}80^\circ$,
$r=1\textrm{--}2$ kpc) and the observed fragment of the Perseus
arm ($l=100\textrm{--}140^\circ$, $r=1.5\textrm{--}3.0$ kpc) can
be associated with some of these model spurs. Such spurs usually
have a larger pitch angle than  large-scale patterns like
segments of elliptical rings or global spiral arms. The nature of
the fragmentation is not clear, it can be of purely
hydro-dynamical origin (Dobbs \& Bonnell 2006) or be associated
with slow modes forming in the stellar population of the disk
(Rautiainen \& Salo 1999, 2000).

\acknowledgements We thank H. Salo for sharing his N-body code.
This work was supported in part by the Russian Foundation for
Basic Research (project nos.~12\mbox{-}02\mbox{-}00827,
13\mbox{-}02\mbox{-}00203,  14\mbox{-}02\mbox{-}00472).
Calculations of the  light-curve parameters and gamma-velocities
of Cepheids  were supported by Russian Scientific Foundation
grant no. 14-22-00041.

Funding for SDSS-III has been provided by the Alfred P. Sloan
Foundation, the Participating Institutions, the National Science
Foundation, and the U.S. Department of Energy Office of Science.
The SDSS-III web site is http://www.sdss3.org/.

SDSS-III is managed by the Astrophysical Research Consortium for
the Participating Institutions of the SDSS-III Collaboration
including the University of Arizona, the Brazilian Participation
Group, Brookhaven National Laboratory, Carnegie Mellon
University, University of Florida, the French Participation
Group, the German Participation Group, Harvard University, the
Instituto de Astrofisica de Canarias, the Michigan State/Notre
Dame/JINA Participation Group, Johns Hopkins University, Lawrence
Berkeley National Laboratory, Max Planck Institute for
Astrophysics, Max Planck Institute for Extraterrestrial Physics,
New Mexico State University, New York University, Ohio State
University, Pennsylvania State University, University of
Portsmouth, Princeton University, the Spanish Participation
Group, University of Tokyo, University of Utah, Vanderbilt
University, University of Virginia, University of Washington, and
Yale University.

\end{document}